\documentstyle[aps]{revtex} 
\begin{document}
\draft
\title{
Parametric resonance at the critical temperature in high energy heavy ion collisions
}
\author{Masamichi Ishihara
\thanks{Electronic address: m\_isihar@koriyama-kgc.ac.jp}
}
\address{Department of Human Life Science , Koriyama Women's University \\
Kaisei 3-5-2, Koriyama, Fukushima, 963-8503, Japan 
}
\date{\today}
\maketitle
\begin{abstract}
Parametric resonance in soft modes at the critical temperature ($T_{c}$)
in high energy heavy ion collisions is studied in the case 
when the temperature ($T$) of the system is almost constant for a long time.
By dividing the fields into three parts, zero mode (condensate), soft modes
and hard modes and assuming that the hard modes are in thermal equilibrium, 
we derive the equation of motion for soft modes at $T=T_{c}$. 
Enhanced modes are extracted by comparing with the Mathieu equation 
for the condensate oscillating along the sigma axis at $T=T_{c}$. 
It is found that the soft mode of $\pi$ fields at about 174 MeV is enhanced. 
\end{abstract}
\pacs{25.75.Dw, 25.75.--q, 11.30.Rd}

\section{Introduction}
It is expected that a new phase of matter will be formed in high energy heavy ion collisions
at RHIC and LHC.
It is called 'quark-gluon plasma' (QGP) in which the chiral symmetry restoration occurs. 
Many signals of the chiral symmetry restoration have been proposed, 
but there is no decisive one.
One of the proposed signals is the disoriented chiral condensate (DCC) 
which is a misalignment phenomena in the chiral space.
The time development of the chiral condensate has been studied in terms of DCC  
and the possibility of soft mode enhancement by parametric resonance was suggested
\cite{Muller}. 
Parametric resonance in the last stage of high energy heavy ion collisions 
was discussed \cite{Muller,Kaiser,Hiro-Oka,Dumitru,Ishihara4,Sornborger,Maedan}
and amplified modes were extracted. 
A parametric resonance is expected even in the chiral phase transition 
because the oscillation of the chiral condensate 
(momentum $k = 0$) may amplify nonzero ($k \neq 0$) modes. 
The motion of the condensate must be investigated in order to reveal this phenomena.
As indicated in \cite{Ishihara23}, 
the motion of the chiral condensate 
in high energy heavy ion collisions is expected to be almost along  
the sigma axis in the linear $\sigma$ model.
The oscillation of the condensate is just that of the sigma condensate. 
There may be a chance of the amplification of the fields by parametric resonance 
if the sigma field oscillates for a long time.

The study of parametric resonance at finite temperature 
in chiral phase transition has never been examined, 
while that at zero temperature 
\cite{Muller,Kaiser,Hiro-Oka,Dumitru,Ishihara4,Sornborger,Maedan}
has been performed.
Parametric resonance may occur even when the temperature is not zero
if the condensates moves periodically like a sine function.
The motion of the condensates can be described by the equation of motion 
with the effective potential if the system is not far from thermal equilibrium.
The effective potential depends on the temperature of the system 
which is a function of (proper) time. 
If the temperature is constant for a long time, 
the condensate will oscillate as it oscillates at zero temperature.   
The periodicity of the motion of the condensate depends on the effective potential at 
finite temperature. 
It requires that the temperature must be almost constant in the period of the oscillation 
of the condensate.

It is expected that the temperature may be  almost constant  as a function of time 
at the critical temperature ($T_{c}$) 
because of the large difference of entropy density between QGP and hadron phases.
Especially, it may take some 10 fm for the phase transition to finish 
if the phase transition from quarks and gluons to hadrons is of the first order \cite{1st}. 
The similar time dependence of the temperature may also occur in 
the second order chiral phase transition of a high energy heavy ion  collision.
The parametric resonance by the oscillation of the sigma condensate 
will occur at/near $T_{c}$ in such a case.
The similar time dependence of the temperature is also expected 
when the cooling of the system is slow. 
If the period of one dimensional scaling \cite{Bjorken} is long enough, 
the temperature is only slowly decreasing 
as a function of time in the last stage of the expansion.

In this paper, we discuss the possibility of parametric resonance at/near $T_{c}$ 
in the framework of the linear $\sigma$ model 
and extract the amplified modes assuming that 
the temperature is constant at $T_{c}$ for a long time.
The paper is organized as follows. 
In Sec. \ref{sec:Equation_of_motion_for_soft_modes}, 
the equation of motion is derived in the case when the temperature is constant at $T=T_{c}$.
In Sec. \ref{sec:Amplified_modes_near},
the amplified modes and the time scale of the amplification
are extracted by comparing with the Mathieu equation.
The time scale is explicitly shown by solving the Mathieu equation numerically.
Section \ref{sec:conclusions} is assigned for conclusions.

\section{Equation of motion for soft modes}
\label{sec:Equation_of_motion_for_soft_modes}

The linear $\sigma$ model is a useful tool 
to describe the motion of the condensate and soft modes
below/near the critical temperature. 
The Lagrangian  is
\begin{equation}
{\cal L} =  \frac{1}{2} \partial_{\mu} \phi \partial^{\mu} \phi 
- \frac{\lambda}{4} \left( \phi^{2} - v^{2} \right)^{2} + H \phi_{0}, 
\label{eqn:Lagrangian}
\end{equation}
where $\phi = (\phi_{0}, \phi_{1}, \phi_{2}, \phi_{3}) = (\sigma,\vec{\pi})$, 
${\displaystyle \phi^{2} = \sum_{j=0}^{3} \phi_{j}^{2}}$ 
and $H$ is the explicit symmetry breaking term.
The field $\phi$ is divided into three parts, 
zero mode ($\phi_{jc}$ , the condensation ), 
soft modes ($\phi_{js}$) and hard modes ($\phi_{jh}$):
\begin{equation}
\phi_{j} = \phi_{jc} + \phi_{js} + \phi_{jh}.
\label{eqn:devide}
\end{equation}
The bracket, $\langle {\cal O} \rangle$, is used to describe 
${\cal O}$ averaged over the hard modes.
We apply the free particle approximation for the hard modes.
Consequently, 
$\langle \phi_{jh}  \rangle$ and $\langle \phi_{jh}^{3}  \rangle$ 
are zero if the distribution of hard modes is thermal.
Substituting eq.(\ref{eqn:devide}) into eq.(\ref{eqn:Lagrangian}) 
and taking the thermal average of hard modes, 
we obtain the following effective Lagrangian:
{
\setcounter{enumi}{\value{equation}}
\addtocounter{enumi}{1}
\setcounter{equation}{0}
\renewcommand{\theequation}{\theenumi\alph{equation}}
\begin{eqnarray}
\langle {\cal L} \rangle &=& \langle {\cal L}_{K} \rangle  +  \langle {\cal L}_{V} \rangle ,
\\
\langle {\cal L}_{K} \rangle &=& 
  \frac{1}{2} \partial_{\mu} \phi_{c} \partial^{\mu} \phi_{c} 
+ \frac{1}{2} \partial_{\mu} \phi_{s} \partial^{\mu} \phi_{s}
+ \partial_{\mu} \phi_{c} \partial^{\mu} \phi_{s}
+ \frac{1}{2} \langle \partial_{\mu} \phi_{h} \partial^{\mu} \phi_{h} \rangle ,
\\
\langle {\cal L}_{V} \rangle &=& 
-\frac{\lambda}{4} \left(\phi_{c}^{2} + \phi_{s}^{2} + 2 \phi_{c} \cdot \phi_{s} - v^{2} \right)^{2}  
+ H \left(  \phi_{0c} + \phi_{0s} \right) 
\nonumber \\ && 
- \frac{\lambda}{2} \langle \phi_{h}^{2}  \rangle 
  \left(\phi_{c}^{2} + \phi_{s}^{2} + 2 \phi_{c} \cdot \phi_{s} - v^{2} \right)
- \frac{\lambda}{4} \langle \left( \phi_{h}^{2}\right)^{2}  \rangle 
- \lambda \sum_{j=0}^{3} \left( \phi_{j c} + \phi_{j s} \right)^{2} \langle \phi_{j h}^{2}  \rangle ,
\end{eqnarray}
where dot implies the inner product defined by 
${\displaystyle  \phi_{c} \cdot \phi_{s} = \sum_{j=0}^{3} \phi_{j c}  \phi_{j s}}$. 
\setcounter{equation}{\value{enumi}}
}

If $\langle {\cal O} \rangle$ terms have no $\phi_{c}$ and $\partial \phi_{c}$ dependence,
Euler-Lagrange equation for $\phi_{j c}$ obtained from $\langle {\cal L} \rangle$ is 

\begin{eqnarray}
&& 
\Box \phi_{j c} + \Box \phi_{j s} 
+ \lambda \left( \phi_{c}^{2} + \phi_{s}^{2} + 2 \phi_{c} \cdot \phi_{s} - v^{2} \right)
  \left( \phi_{j c} + \phi_{j s } \right) - H \delta_{j 0} 
\nonumber \\ && \hspace{1cm}
+ \lambda \langle \phi_{h}^{2} \rangle \left( \phi_{j c} + \phi_{j s} \right) 
+ 2 \lambda  \langle \phi_{j h}^{2} \rangle \left( \phi_{j c} + \phi_{j s} \right) 
= 0.
\label{eqm:basic}
\end{eqnarray}
An equation of the same form for $\phi_{j s}$ is obtained 
if $\langle {\cal O} \rangle$ terms have no $\phi_{s}$ and $\partial \phi_{s}$ .
Here it is assumed that 
$\langle \phi_{j h}^{2} \rangle \stackrel{\rm def}{=} {\cal F}(T)$ is $j$-independent.
The meaning of this assumption becomes apparent 
when the concrete expression of ${\cal F}(T)$ is obtained in the next section.
We introduce the effective potential defined by 
\begin{equation}
V(\phi,\phi_{0};T) = \frac{\lambda}{4} \left( \phi_{c}^{2} + 6 {\cal F}(T) - v^{2} \right)^{2} - H \phi_{0c} .
\end{equation}
Note that the order of the phase transition described by this potential is second.

Since we are interested in the amplification of the soft mode with small amplitude, 
we first consider eq.(\ref{eqm:basic}) with $\phi_{j s} = 0$ for all $j$.
This is the zeroth order equation of soft modes: 
{
\setcounter{enumi}{\value{equation}}
\addtocounter{enumi}{1}
\setcounter{equation}{0}
\renewcommand{\theequation}{\theenumi\alph{equation}}
\begin{equation}
\Box \phi_{j c} + \partial V/ \partial \phi_{j c} = 0  .
\label{eqn:eq_for_cond}
\end{equation}
The equation for soft modes with small amplitudes
is obtained by substituting eq.(\ref{eqn:eq_for_cond}) 
into eq.(\ref{eqm:basic}):
\begin{equation}
\Box \phi_{j s} 
+ \lambda \left( \phi_{c}^{2} + \phi_{s}^{2} + 2 \phi_{c} \cdot \phi_{s} + 6 {\cal F}(T)  - v^{2} \right) \phi_{j s} 
+ \lambda \left(2 \phi_{c} \cdot \phi_{s} + \phi_{s}^{2} \right) \phi_{j c} = 0 .
\label{eqn:eq_for_soft}
\end{equation}
\setcounter{equation}{\value{enumi}}
Eq.(\ref{eqn:eq_for_soft}) is the equation of motion 
for $\phi_{s}$ with the back ground field, $\phi_{c}$.

We would like to calculate the critical temperature.
In a realistic case, the critical temperature cannot be defined 
exactly since  $H$ is not zero. 
Nevertheless, one can estimate the critical temperature $T_{c}$ 
by requiring that  one false minimum disappears at $T=T_{c}$: 
}
{
\setcounter{enumi}{\value{equation}}
\addtocounter{enumi}{1}
\setcounter{equation}{0}
\renewcommand{\theequation}{\theenumi\alph{equation}}
\begin{eqnarray}
&& \left( \bar{\phi}_{0c}^{2} + 6 {\cal F}(T_{c}) - v^{2} \right) \bar{\phi}_{0c} - H / \lambda = 0, 
\label{cond:min} 
\\
&& \bar{\phi}_{nc} = 0 \hspace{0.25cm} (n=1,2,3) ,
\end{eqnarray}
\setcounter{equation}{\value{enumi}}
where $\bar{\phi}_{j c}$ is the condensation 
which is the expectation value of the field $\phi_{j}$ at 
the minimum of the effective potential at $T = T_{c}$ [eq.(\ref{cond:min})]. 
$T_c$ can be obtained by the condition that 
two solutions of eq.(\ref{cond:min}) are the same:
}
\begin{equation}
{\cal F}(T_{c}) =  \frac{v^{2}}{6} - \frac{1}{8} \left( \frac{4H}{\lambda} \right)^{2/3} .
\label{eqn:findTc}
\end{equation}
We introduce the fluctuation fields $\delta \phi_{j c}$ 
defined by 
\begin{equation}
\delta \phi_{j c} = \phi_{j c} - \bar{\phi}_{j c}.
\label{def:flac}
\end{equation}
Substituting eq.(\ref{def:flac}) into eq.(\ref{eqn:eq_for_cond}) and 
using the condition, eq.(\ref{cond:min}),  
we have
\begin{equation}
\Box \left( \delta \phi_{j c} \right) 
+ \lambda \left[ \left( \delta \phi \right)^{2} + 2 \left( \delta \phi \right) \cdot \bar{\phi} \right] 
  \left[ \bar{\phi}_{0 c} +  \left( \delta \phi_{j c} \right) \right] 
+ \left( H/\bar{\phi}_{0c} \right) \left( \delta \phi_{j c} \right) = 0 .
\end{equation}
Ignoring $O(\delta^{2})$ and higher terms, we obtain 
\begin{equation}
\left( \Box + m_{j}^{2}\right) \left( \delta \phi_{j c} \right) = 0 ,
\label{eqn:condensate}
\end{equation}
where 
$m_{0}^{2} \stackrel{\rm def}{=} \left( H / \bar{\phi}_{0c} \right) + 2 \lambda \bar{\phi}_{0c}^{2}$
and 
$m_{n}^{2} \stackrel{\rm def}{=} \left( H / \bar{\phi}_{0c} \right)$ $(n=1,2,3)$. 
It is easily found  from eq.(\ref{eqn:condensate}) that 
the condensate oscillates around the minimum of the potential at $T_c$. 
In the same way, eq.(\ref{eqn:eq_for_soft}) becomes 
\begin{eqnarray}
&& 
\Box \phi_{j s} + \left( H / \bar{\phi}_{0c} \right) \phi_{j s} 
+ \lambda \left( \phi_{s}^{2} + 2 \bar{\phi}_{c} \cdot \phi_{s} \right) \phi_{js} 
+ 2 \lambda \left[ \bar{\phi}_{c} \cdot \left( \delta \phi_{c} \right) 
  + \phi_{s} \cdot \left( \delta \phi_{c} \right) \right] \phi_{j s} 
\nonumber \\ && \hspace{1cm} 
+ \lambda \left( \delta \phi_{c} \right)^{2} \phi_{j s}
+ \lambda \left( 2 \bar{\phi}_{c} \cdot \phi_{s} + \phi_{s}^{2} 
  + 2 \left( \delta \phi_{c} \right) \cdot \phi_{s} \right) \phi_{j c} = 0 .
\label{eqn:softmodes}
\end{eqnarray}
Since we are interested in the amplification by the oscillation of the condensate, 
we consider the small $\phi_{s}$ and discard $O(\phi_{s}^{2})$ and $O(\phi_{s}^{3})$ terms.
The equation for $\phi_{j s}$ [eq.(\ref{eqn:softmodes})] in such a case is 
\begin{equation}
\Box \phi_{j s} + \alpha_{j} \phi_{j s} 
+ \sum_{\stackrel{i = 0}{(i \neq j)}}^{3} \beta_{j i} \phi_{i s} = 0 ,
\label{eqn:soft_non_diag}
\end{equation}
where 
{
\setcounter{enumi}{\value{equation}}
\addtocounter{enumi}{1}
\setcounter{equation}{0}
\renewcommand{\theequation}{\theenumi\alph{equation}}
\begin{eqnarray}
\alpha_{j} &\stackrel{\rm def}{=}&  
H / \bar{\phi}_{0c} + 2 \lambda \left[ 
\bar{\phi}_{c} \cdot \left( \delta \phi_{c} \right) + \left( \bar{\phi}_{j c} \right)^{2} 
+ 2 \bar{\phi}_{j c} \left( \delta \phi_{j c} \right)
\right] , 
\\
\beta_{j i}  &\stackrel{\rm def}{=}& 
2 \lambda \left[
\bar{\phi}_{j c} \bar{\phi}_{i c} + \bar{\phi}_{j c} \left( \delta \phi_{i c} \right)
+  \bar{\phi}_{i c} \left( \delta \phi_{j c} \right) 
\right]  
\equiv \beta_{i j}
\hspace{1cm} (j \neq i)  ,
\end{eqnarray}
and $O(\delta^{2})$ terms have been ignored.  
Since $\bar{\phi}_{0c} \neq 0$ and $\bar{\phi}_{nc} = 0 (n=1,2,3)$ 
in the linear $\sigma$ model, 
the coefficients $\alpha_{j}$ and $\beta_{j i}$ have the following relations:
\setcounter{equation}{\value{enumi}}
}
\begin{equation}
\alpha_{1} = \alpha_{2} = \alpha_{3} , \hspace{1cm}
\beta_{12} = \beta_{13} = \beta_{21} = \beta_{23} = \beta_{31} = \beta_{32} = 0 .
\end{equation}

As stated in Introduction, the condensate moves and oscillates almost along the sigma axis. 
Then, $\delta \phi_{nc} \sim  0$ for $n=1,2,3$. Consequently all $\beta$ is zero.
In such a case, eq.(\ref{eqn:soft_non_diag}) is diagonalized:
{
\setcounter{enumi}{\value{equation}}
\addtocounter{enumi}{1}
\setcounter{equation}{0}
\renewcommand{\theequation}{\theenumi\alph{equation}}
\begin{eqnarray}
&& 
\Box \phi_{0s} 
+ \left[ 
m_{0}^{2} 
+ 6 \lambda \bar{\phi}_{0c} \left( \delta \phi_{0c} \right) 
\right] \phi_{0s} = 0 ,
\label{eqn:sigma_soft}
\\ && 
\Box  \phi_{ns} 
+ \left[ 
m_{n}^{2} 
+  2 \lambda \bar{\phi}_{0c} \left( \delta \phi_{0c} \right) 
\right]  \phi_{ns} = 0 .
\label{eqn:pi_soft}
\end{eqnarray}
\setcounter{equation}{\value{enumi}}
The solution of eq.(\ref{eqn:condensate}) for $j=0$ is 
}
\begin{equation}
\delta \phi_{0c} = - B \cos \left( m_{0} t + \theta \right) ,
\label{eqn:sol_of_cond}
\end{equation}
where $B$ and $\theta$ in eq.(\ref{eqn:sol_of_cond}) are determined by 
the initial condition of $\delta \phi_{0c}$.  
This solution, eq.(\ref{eqn:sol_of_cond}), is substituted into 
eqs.(\ref{eqn:sigma_soft}),(\ref{eqn:pi_soft}) and 
the transformation from $t$ to $\xi \stackrel{\rm def}{=} ( m_{0}t + \theta) / 2$ 
is applied. 
The equations of motion now become:
{
\setcounter{enumi}{\value{equation}}
\addtocounter{enumi}{1}
\setcounter{equation}{0}
\renewcommand{\theequation}{\theenumi\alph{equation}}
\begin{eqnarray}
&& 
\left\{
\frac{d^{2}}{d\xi^{2}} + \frac{4}{m_{0}^{2}} \left(\vec{k}^{2} + m_{0}^{2}\right) 
- \frac{24 \lambda \bar{\phi}_{0c} B}{m_{0}^{2}} \cos (2 \xi) 
\right\} \phi_{0s}(\xi ;\vec{k})  = 0 ,
\label{eqn:final_sigma}
\\ && 
\left\{
\frac{d^{2}}{d \xi^{2}} + \frac{4}{m_{0}^{2}} \left(\vec{k}^{2} + m_{n}^{2}\right) 
- \frac{8 \lambda \bar{\phi}_{0c} B}{m_{0}^{2}} \cos (2 \xi) 
\right\} \phi_{ns}(\xi;\vec{k})  = 0 ,
\label{eqn:final_pi}
\end{eqnarray}
where $\phi_{0s}(\xi;\vec{k})$ and $\phi_{ns}(\xi;\vec{k})$ 
are the Fourier transformation 
of $\phi_{0s}(\xi;\vec{x})$ and $\phi_{ns}(\xi;\vec{x})$, respectively.
\setcounter{equation}{\value{enumi}}
}

\section{Amplified modes} 
\label{sec:Amplified_modes_near}
Eqs.(\ref{eqn:final_sigma}),(\ref{eqn:final_pi}) are just Mathieu equation.
To investigate the amplified modes, we define the following quantities:
\begin{equation}
A_{\sigma} = \frac{4}{m_{0}^{2}} \left(\vec{k}^{2} + m_{0}^{2}\right),
\hspace{1cm}
A_{\pi} = \frac{4}{m_{0}^{2}} \left(\vec{k}^{2} + m_{n}^{2}\right) .
\end{equation}
The amplified modes are obtained from the above coefficients by the help of 
the knowledge of Mathieu equation.
The amplified modes of $\sigma$ field 
for nonzero modes ($k \neq 0$) corresponds to $A_{\sigma} = 9,16,\cdots$ 
because $A_{\sigma} > 4$ apparently,
while that of $\pi$ field correspond to $A_{\pi} = 1,4,9,16,\cdots$. 
($A_{\pi}=1$ is not satisfied for some parameters of the linear $\sigma$ model.) 
Then, the masses of $\sigma$ and $\pi$ fields are needed to determine such modes.
The condensate ($\bar{\phi}_{0c}$) at $T_{c}$ is $(4 H/\lambda)^{1/3}$ which is 
easily found from the definition of $T_{c}$. 
Therefore, the masses are obtained if $\lambda$ and $H$ are given.
The amplified modes corresponding to $A_{\sigma}, A_{\pi} = 1,4,9,16$ are 
shown in Table \ref{tbl:modes} for $\lambda =20$ and $H^{1/3} = 119$ MeV
which generate $m_{\pi}$ = 135 MeV, $m_{\sigma}$ = 600 MeV and $f_{\pi}=92.5$ MeV
at $T = 0$ for $v = 87.4$ MeV.

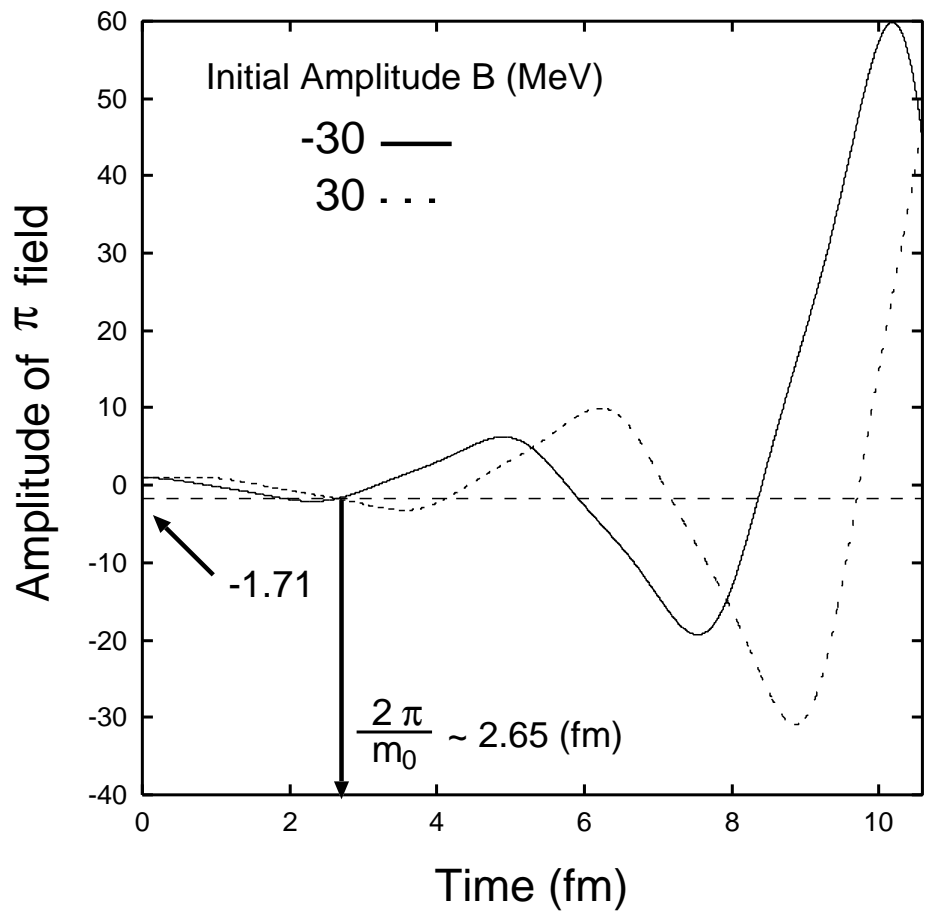
\begin{figure}
\caption{
The solution $w(t)$ of the Mathieu equation for $\pi$ fields 
with $A_{\pi} = 1$ with the initial amplitude $B= \mp 30$ MeV and 
$w(0)=1$, $dw/dt(0) =0$.
Thick line and dotted line are for $B=-30$ and $30$ MeV respectively.
Dashed line is for $-1.71$ which is the value of $\cos(\nu \xi)$ at $\xi = \pi$.
The arrow with the string '$2\pi/m_{0}$' shows the time corresponding to $\xi = \pi$.
}
\label{fig:numerical_pi}
\end{figure}

The amplification of $\pi$ fields is determined by $A_{\pi}$ and
the factor ($2Q_{\pi}$) in the presence of cosine in the Mathieu equation.
This factor is given by 
\begin{equation}
2Q_{\pi} = \frac{8 \lambda \bar{\phi}_{0c} B}{m_{0}^{2}},
\end{equation}
where $B$ is the amplitude of zero mode introduced in eq.(\ref{eqn:sol_of_cond}).
Its numerical value 
for the previous parameters of the linear $\sigma$ model is about
$ 0.051 \ ({\rm MeV}^{-1}) \times B $.
For example, $2Q_{\pi}$ is $-1.53$ for $B=-30$ MeV.
Since the solution , $w(\xi)$, of the Mathieu equation for $w(0)=1, dw/d\xi (0) = 0$ 
has the property 
$w(\xi+\pi) =  e^{i \nu \pi} w(\xi)$ where $\nu$ is called 'the characteristic exponent'
which has the relation:
\begin{equation} 
\cos(\nu \pi) = w(\pi).
\end{equation} 
The time period, $\pi$, in $\xi$ corresponds to about 2.65 fm in $t$. 
It is obtained that
$ \left| e^{i \nu \pi} \right| \sim 3.1$ \cite{Abramowitz}  
for $A_{\pi}=1$ and $ -2Q_{\pi}/A_{\pi} = 1.53$.
The numerical solutions for $B = \mp 30$ MeV for  $w(0)=1, dw/dt (0) = 0$ 
are shown in Fig.\ref{fig:numerical_pi}. 
$\omega(\xi=\pi)$ is about $-1.71$, 
which corresponds to $ \left| e^{i \nu \pi} \right| \sim 3.1 $. 
It is found that the fields $\pi$ are strongly amplified in 
a few (5 or more) fm. 

There are several amplified modes in Mathieu equation,
eqs.(\ref{eqn:final_sigma}),(\ref{eqn:final_pi}).
However, modes larger than $k_{\Lambda}$ cannot be amplified modes actually
because we assume that hard modes are in thermal. 
Therefore, the cutoff ($k_{\Lambda}$) between soft and hard modes is important 
in order to know whether the amplified modes extracted 
from eqs.(\ref{eqn:final_sigma}),(\ref{eqn:final_pi}) 
belong to soft modes or not.
In this paper, $k_{\Lambda}$ is determined as follows. 
Since ${\cal F}(T)$ is a function of $k_{\Lambda}$, 
the latter is obtained from eq.(\ref{eqn:findTc}) in which $T_{c}$ 
is related to $k_{\Lambda}$. 
Namely $k_{\Lambda}$ is fixed if $T_{c}$  is given. 
The density operator of $j$ field for hard modes [$\rho^{j}_{h}(T)$] 
is assumed as follows:
\begin{equation} 
\rho^{j}_{h}(T) = 
\exp \left( 
- T^{-1} \sum_{ \left| \vec{k} \right| \ge k_{\Lambda}} 
\omega_{j} a_{j}^{\dag}(\vec{k}) a_{j}(\vec{k}) 
\right)
/ {\rm Tr} \left[
\exp \left( 
- T^{-1} \sum_{ \left| \vec{k} \right| \ge k_{\Lambda}} 
\omega_{j} a_{j}^{\dag}(\vec{k}) a_{j}(\vec{k}) 
\right)
\right],
\end{equation} 
where $\omega^{2}_{j} = \vec{k}^{2} + m_{j}^{2}$.
We find:
\begin{equation}
{\cal F}(T) \equiv
\langle \phi_{j h}^{2}(x) \rangle = 
{\rm Tr} \left( \rho_{h}^{j}(T) \phi_{j h}^{2}(x) \right) = 
\frac{1}{2V} \sum_{ \left| \vec{k} \right| \ge k_{\Lambda}} \omega_{j}^{-1}
+ 
\frac{1}{V} \sum_{\left| \vec{k} \right| \ge k_{\Lambda}}
\frac{1}{\left[ \exp(\omega_{j}(\vec{k})/T) - 1 \right] \omega_{j}(\vec{k})}.
\label{eqn:phi2}
\end{equation}
The first term is the vacuum contribution and the second is the thermal one.
The vacuum contribution 
can be  
discarded in the following calculation
since it is the well-known infinity which is removed by the redefinition of the energy 
including the contribution from the soft modes.
(Subtraction of the vacuum energy)
Eq.(\ref{eqn:findTc}) is rewritten by
the massless particle approximation and the integration of angle:
\begin{equation} 
\frac{1}{\pi^{2}} \int_{k_{\Lambda}/T_{c}}^{\infty} du \frac{u}{\exp(u) - 1} 
= \frac{1}{3} \left( \frac{v}{T_{c}} \right)^{2} 
- \frac{1}{4 T_{c}^{2}} \left( \frac{4H}{\lambda} \right)^{2/3} .
\label{eqn:kLambda}
\end{equation} 
This approximation is valid near $T_{c}$
because the masses of $\sigma$ and $\pi$ mesons become small.
These masses are zero at $T=T_{c}$ in the chiral limit.
Note that 
$\langle {\cal O} \rangle$ terms have no dependence on 
$\partial \phi_{0}, \phi_{0}, \partial \phi_{s}$ and $\phi_{s}$ 
because
${\cal F}(T)$ is independent of $j$ in the massless particle approximation.

Table \ref{tbl:khard} shows $k_{\Lambda}$ 
obtained by solving eq.(\ref{eqn:kLambda}) for various $T_{c}$ 
with $\lambda = 20$, $H^{1/3}=119$ MeV and $v =87.4$ MeV.
$T_{c}$ is about 123 MeV in the chiral limit. 
It 
has been estimated from Lattice QCD and some other methods.
It is between 140 MeV and 190 MeV in Lattice QCD \cite{QCD0,QCD}.  
Since $k_{\Lambda}$ is above 190 MeV except for $T_{c} \le  130$ MeV, 
we conclude that $k_{\pi} \sim 174$ MeV is an amplified mode at least
when $T_{c}$ is adjusted above 130 MeV. 
Other modes are irrelevant 
since these modes belong to hard modes.
( $k_{\pi} \sim 440.1$ MeV may be a soft mode in this sense.)

The amplified mode of $\pi$ field is not directly related to the observed one 
since the mass of $\pi$ meson at $T \neq 0$ is different from  that at $T = 0$.
Then we must know the effect of the difference of two masses.
In quantum field theory, the relation between these two masses can be 
described by Bogoliubov transformation. 
we can estimate the effect by evaluating the coefficients of this transformation.
The amplified mode is not changed essentially
because 
the difference of the masses
between at zero temperature and at the critical temperature is small enough
for $\pi$ fields in this model. 
The peak coming from the parametric resonance at $T=T_{c}$ will be found 
near $k_{\pi} \sim 174$ MeV if the peak is not smeared out 
by scattering, absorption and so on.

\section{Conclusions}
\label{sec:conclusions}

The parametric resonance at the critical temperature
in high energy heavy ion collisions is studied
in the case when the temperature of the system 
is constant at $T_{c}$ for a long time.
We consider the case in which 
the condensate oscillates along the sigma axis
at the critical temperature of the second order phase transition  
in the framework of the linear $\sigma$ model.

The enhancement of soft mode at about 174 MeV in $\pi$ field is found 
at various $T_{c}$  above 140 MeV.
Other modes are irrelevant 
as amplified modes 
because these modes belong to the hard modes which are assumed to be 
in thermal equilibrium in the present study.
On the contrary, there is no enhanced soft mode in the $\sigma$ field.  
The amplified mode (174 MeV) at $T = T_{c}$ is softer than 
that (for example, about 265 MeV in one dimensional expansion case \cite{Ishihara4}) 
at $T=0$ in $\pi$ field 
because the difference between sigma and pion masses at $T=T_{c}$ is smaller
than that at $T=0$.
It takes a short time for the soft modes to be amplified. 
This implies that the amplification of the soft mode by 
the parametric resonance at $T_{c}$ is possible 
in real collisions at high energies. 
The enhancement by the parametric resonance in the present study may be observed experimentally
if the smearing effects are weak enough. 

It has been pointed out that there may be the parametric resonance at zero temperature
in the last stage of heavy ion collisions.
Then, the pion momentum distribution caused by the parametric resonance 
at finite temperatures may be the initial distribution of the subsequent parametric 
resonance at zero temperature. 
If so, some peaks which correspond to the enhanced modes may appear 
in the final momentum distribution.

The explicit symmetry breaking term ($H \phi_{0}$)
plays important roles for the parametric resonance at $T=T_{c}$ 
because this term makes the condensate $\bar{\phi}_{0c}$ be nonzero 
 and generates masses at $T=T_{c}$.
The condensates ($\bar{\phi}_{jc}$) and the masses ($m_{j}$) at $T_{c}$ 
are exactly zero in the chiral limit ($H=0$).
Note that 
eqs.(\ref{eqn:sigma_soft}),(\ref{eqn:pi_soft}) are valid 
when $O\left(\phi_{s}^{2}\right)$  and $O\left(\phi_{s}^{3}\right)$ terms 
in eq.(\ref{eqn:softmodes}) are negligible.
The resonance structure may change in small $H$ cases. 

We have not shown the strength 
of the peak in the present study. 
If we would like to obtain it, we must know the time interval in which 
the temperature is (almost) constant and consider the evolution of QGP.
The non-linear effects and back reaction are ignored also in this investigation.
These problems will be answered in the future studies.

\begin{acknowledgments}
I would like to thank F. Takagi for a number of helpful suggestions.
\end{acknowledgments}


\clearpage
\begin{table*}
\caption{
Amplified modes in $\sigma$ and $\pi$ fields for various $A_{\sigma}$ and $A_{\pi}$.
}
\label{tbl:modes}
\end{table*}

\begin{table*}
\caption{
The relation between $T_{c}$ and $k_{\Lambda}$. 
}
\label{tbl:khard}
\end{table*}


\clearpage

\begin{center}
\begin{tabular}{|c|c|c|}
\hline
$A_{\sigma}$ ,$A_{\pi}$  & $k_{\sigma}$ (MeV) & $k_{\pi}$ (MeV) \\ 
\hline
1 & --    & 174.0 \\
4 & 0     & 440.1 \\
9 & 521.9 & 682.7 \\
16& 808.5 & 920.5 \\
\hline
\end{tabular}
\end{center}
\bigskip
\begin{center}
Table.\ref{tbl:modes}
\end{center}

\bigskip \bigskip \bigskip

\begin{center}
\begin{tabular}{|c|c|}
\hline
$T_{c}$ (MeV)  &  $k_{\Lambda}$ (MeV) \\ 
\hline
89.5 & 0 \\ 
130 & 150.5 \\
140 & 192.3 \\
150 & 235.7 \\
160 & 280.5 \\
170 & 326.8 \\ 
180 & 379.3 \\
\hline
\end{tabular}
\end{center}
\bigskip
\begin{center}
Table.\ref{tbl:khard}
\end{center}

\clearpage

\end{document}